\documentstyle[preprint,aps]{revtex}
\begin{document}
\draft
\preprint{HU-TFT-95-47, DUKE-TH-95-96, LBL-37642}

\title{Screening of initial parton production in ultrarelativistic heavy-ion
collisions}
\author{K. J. Eskola}
\address{Laboratory of High Energy Physics and
Research Institute for Theoretical Physics, \hspace*{5mm}
P.O. Box 9, FIN-00014 University of Helsinki, Finland}
\author{B. M\"uller}
\address{Department of Physics, Duke University, Durham, NC 27708-0305, U.S.A.}
\author{Xin-Nian Wang}
\address{Nuclear Science Division, MS 70A-3307,
         Lawrence Berkeley Laboratory, \\
         University of California, Berkeley, CA 94720, U.S.A.}
\date{August, 1995}
\maketitle

\begin{abstract}

\baselineskip=18pt

Screening of initial parton production due to the presence of
on-shell partons in high-energy heavy-ion collisions is
discussed. It is shown that the divergent cross sections in the
calculation of parton production can be regulated
self-consistently without an {\it ad hoc} cut-off, and that
the resultant parton production and transverse energy
production rate are finite. Consequences on the energy density
estimates are discussed.

\end{abstract}
\pacs{PACS numbers:12.38.Mh,25.75+r,12.38.Bx}

\baselineskip=16pt

        Hard and semihard parton scatterings have been shown to play
an important role in both high energy $pp(\bar{p})$ and
ultrarelativistic heavy ion collisions.
In $pp(\bar p)$ collisions they are believed to be responsible
for the  rapid growth of the inelastic and total cross sections
\cite{GH,XNW}, and in $AA$ collisions they are expected to produce
a large amount of transverse energy in the central rapidity region
\cite{JBAM,KLL,HIJING,PCM}. Because the dominant QCD parton
cross sections are singular in the soft
scattering limit, most model calculations
of minijet production via perturbative QCD (pQCD) have
introduced an  infrared cut-off, $p_0$,
corresponding to the smallest permissible transverse momentum transfer
in a $2\rightarrow2$ parton scattering, to which perturbative
QCD can still be applied.  This cut-off is used in most of the models
to separate perturbative hard processes from nonperturbative soft
interactions which can be modeled, e.g., by strings or flux tubes. Since
there is no distinct boundary between soft and hard physics,
both the hard and soft part of the interaction in this scheme are
very sensitive to the cut-off $p_0$.

Heavy ion collisions
differ from $pp$ collisions in that minijets are produced in large
number, so that a medium of minijets is formed. In the space-time
evolution of a heavy-ion collision  soft particle production  will
be completed after semihard and hard processes. Therefore, the
soft interactions are expected to be
screened by interactions with the semihard quanta (minijets).
We here take a step further by proposing that
the perturbative semihard particle production should also be
screened by the processes which have happened even earlier,
i.e., by the harder processes.
In this paper we will study the influence of the final-state
interactions among minijets. We will
investigate to what extent  minijet production is
screened and  whether
the screening effect can provide an effective self-consistent
regularization of the infrared behavior of minijet production.

        Consider an example of two hard processes as illustrated in
{}Fig.~1. Let us assume jets from the first hard scattering are
produced in the 90 degree direction and have a large transverse
momentum $p_{T1}$. Because of this large transverse momentum,
the interaction point is well localized in the transverse direction
to a distance of $1/p_{T1}$. As these two jets travel in the transverse
direction, they will experience secondary interactions. These
secondary interactions can give rise to many nuclear effects
of hard scatterings, i.e., energy loss and Cronin effects.
Since we are interested in the screening effects caused by these
final state interactions, we will consider the interactions of the
produced hard partons with the propagating parton in another
scattering nearby as shown in Fig.~1. A semiclassical estimate
of the screening requires that different scattering events can
be treated as incoherent. We will show that this condition is
satisfied if the produced partons, which screen
other softer interactions, can be treated as on-shell
particles.

        In a parton model, we treat partons from the nuclei
and out-going partons as on-shell. Therefore, the propagator
in the first hard scattering must be off-shell by $-p_{T1}^2$.
One notices that there is one extra loop in the Feyman
diagram in Fig.~1 as compared to that of two independent
scatterings. The loop-momentum integration will then put
one of the propagators on shell. Since we demand that the second
scattering also has transverse momentum transfer $p_{T2}$,
the propagators in this scattering will also be off-shell
by $-p_{T2}^2$. Therefore, the dominant contribution to the
loop-momentum integration comes from the pole of the propagator
between the two scatterings in which the propagating parton
will be put on-shell. The condition for this contribution to
be dominant is that the distance $\Delta x_{\perp}$ between the
two scatterings must be larger than the interaction range of
the two hard scatterings which are determined by the off-shellness
of the exchanged gluons, i.e.,
$\Delta x_{\perp}>{\rm max}(1/p_{T1},1/p_{T2})$. If this condition
is not satisfied, the propagating parton between two scatterings
cannot be treated as real and consequently one cannot treat the
multiple scatterings as incoherent scatterings. One
can thus consider
\begin{equation}
        \tau_f(p_T)=\frac{a}{p_T}  \label{ftime}
\end{equation}
as the formation time of the produced parton in the mid-rapidity
from the hard or
semihard scatterings after which they can be treated as real
(on-shell) particles and can screen other interactions with
smaller transverse momentum transfer. Here we used a
dimensionless coefficient $a$ of the order unity
to characterize the uncertainty in the formation time.

        Let us visualize high-energy $AA$ collisions
as in Fig.~\ref{fig2}. In this  idealized space-time
picture the incoming nuclei are very thin slabs,
the primary semihard parton-parton collisions all start at $t=0$
and the forming system at central rapidity is longitudinally boost invariant.
In the space-time evolution of the collision the
partons with larger $p_T$ are produced earlier,
as implied by Eq.~\ref{ftime}. These hard partons will
then screen production of partons with smaller $p_T$
later in time and space. For the central region around
$y=0$, we will consider the screening effect of
partons within a unit rapidity interval, $\Delta y=Y=1$.
We assume minijets have a plateau in the rapidity distribution,
which is a good approximation both at LHC and RHIC energies for
our choice of  $Y$. For a fixed $p_T$, partons with
larger rapidities form later in our given frame,
along the proper time hyperbola
as in Fig.~\ref{fig2}. In this picture, the formed
partons can therefore only screen
those scatterings which are in the same rapidity region.

        Following Ref.~\cite{BMW}, we can estimate the static electric
screening mass generated by the produced minijets. The number distribution of
minijets produced in an $AA$ collision at an impact parameter ${\bf b=0}$
can be written as \cite{KLL}
\begin{equation}
        \frac{dN_{AA}}{dp_T^2dy} =
        T_{AA}({\bf 0})\frac{d\sigma_{jet}}{dp_T^2dy},
\label{NAA}
\end{equation}
where $T_{AA}({\bf b})$ is the nuclear overlap function and
\begin{equation}
        \frac{d\sigma_{jet}}{dp_T^2dy} = K\sum_{{ijkl=}\atop{q,\bar q,g}}
        \int dy_2x_1f_{i}(x_1,p_T^2) x_2f_{j}(x_2,p_T^2)
        {d\hat\sigma\over d\hat t}^{ij\rightarrow kl}
        {\hskip-7mm}(\hat s,\hat t,\hat u)
\label{minijets}
\end{equation}
is the minijet cross section. The hats refer to the partonic
sub-processes, $x_i$ is the momentum fraction of the initial state parton
$i$, $p_T$ is the transverse momentum, $y_k$ is the
rapidity of the final state parton $k$, $f_i$ is the parton distribution
function, and $K=2$ is a factor simulating the contribution
from the ${\cal O}(\alpha_s^3)$ terms in the cross section
\cite{EKS,EWinHP}.
In what follows, we will compute the minijet production from
the complete lowest order formula in  Eq.~\ref{minijets} but
will finally treat all the minijets as gluons. This should
again be a good approximation, since gluons clearly dominate
the minijet production at these energies and transverse momenta
\cite{EKR94}.

        To obtain an estimate of the average parton number density in the
central region at a given time $\tau_f$, we divide Eq.~\ref{NAA}
by the approximate volume
$V=\pi R_A^2\Delta z\approx \pi R_A^2\tau_f\Delta y$
of the produced system. Then the screening
mass from Ref.~\cite{BMW} becomes
\begin{equation}
        \mu_D^2(p_T)\approx\frac{3\alpha_{\rm s}}{R_A^2\tau_f(p_T)Y}
        2\arcsin[{\rm tanh}(Y/2)]\int_{p_T}^{\sqrt{s}/2}dk_T
        \frac{dN_{AA}}{dk_T^2dy}\bigg|_{y=0}.
\label{muD2}
\end{equation}

        At this point there are several remarks to be made.
First, as done in Ref.~\cite{BMW}, we have replaced the thermal
(isotropic) Bose-Einstein distribution by the non-isotropic
minijet distribution, and assumed that the screening mass can
still be calculated  from the equilibrium formula given in ~\cite{BMW}.

        Second, note that the lower limit in the $k_T$-integration
in Eq.~\ref{muD2} is $p_T$. In this way, we are considering a
maximal screening effect by assuming that all the quanta with
transverse momenta $k_T\ge p_T=a/\tau$ screen the formation
of partons at transverse momenta $k_T< p_T$.
Also, as explained in the previous paragraphs, only the
quanta within the rapidity window $Y$ are assumed to
contribute to the screening mass. However, from Eq.~\ref{muD2}
we see that as far as the interval $Y$ is small enough, the
screening mass is almost independent of the choice for $Y$,
due to the division by the volume of the system.
Finally, the scale choice for the running coupling is not clear.
Here we just use the $p_T$ of the primary semihard processes.

        As we know from studies of pQCD at finite temperature, perturbation
theory does not generate static screening in the magnetic
part of interactions \cite{WELD82,KAP85}. One normally has to introduce
a nonperturbative magnetic mass to regulate the infrared singularities
in the transverse part of a gluon propagator.
Since the magnetic screening mass in the static limit is not known,
let us proceed in the following phenomenological way
to get an order of magnitude
estimate on the screening effects. We will use the computed
electric screening mass as a regulator for the divergent
$\hat t$- and $\hat u$-channel sub-processes.
We will simply make a replacement
$-\hat t(\hat u)\rightarrow -\hat t(\hat u) + \mu_D^2$ in the
minijet cross section of Eq.~\ref{minijets}.  Since
the $\hat s$-channel processes are not important here,
we do not need to make the above replacement for $\hat s$.
Then we are also safe from the contradiction that the invariant
$\hat s$ is always positive but the screening mass $\mu_D$
is computed for  a spacelike gluon.

        In Fig.~\ref{fig3} we show the screening mass $\mu_D$
and the screened one-jet cross section  as functions of
$p_T$. In the upper panel the results are shown for the LHC energy
(per nucleon pair)
$\sqrt s=5.5$ TeV and in the lower panel for the RHIC energy
$\sqrt s=200$ GeV. We start our numerical computation at sufficiently
high $p_T$ (i.e. at early times $\tau_f$) where  the screening mass
is small enough and the cross section is not screened in any significant
way. At this $p_T$ we then compute the screening mass
$\mu_D^2(p_T)$ from Eq. \ref{muD2}
and with the obtained $\mu_D^2(p_T)$  we compute the screened cross
section at $p_T-\Delta p_T$ (with a sufficiently small step $\Delta p_T$).
{}From the resultant cross section we get the new screening mass at
$p_T-\Delta p_T$ and so on. This iterative procedure gives us
a self-consistent picture of screening: the bigger the cross section becomes
the denser the parton medium gets and the larger a screening mass
is generated, slowing down the rise of the cross section towards
smaller $p_T$. In this way, the medium of produced minijets
regulates the rapid growth of the jet cross section as seen
in Fig.~\ref{fig3}. Finally, at $\mu_D\sim p_T$, the cross section
saturates. From  Fig.~\ref{fig3} we see that for sufficiently
large nuclei and for high collision energies this happens
in the perturbative region
$p_T\gg \Lambda_{QCD}$. To study the sensitivity of the results
to the uncertainty relation (1), we show curves corresponding to
$a=1$ and $a=2$. Note that due to the
self-consistent iteration procedure the resulting difference in $\mu_D$ is
not a trivial multiplicative factor $\sqrt 2$.

        In the computation we have used the MRSA parton distributions
$f_i(x,Q)$ with a scale choice $Q=p_T$ \cite{MRS}.
At this point it is worthwhile to note that the
singular parton distributions are stable against scale evolution,
and  the small-$x$ growth $xf_i\sim x^{-\delta}$
persists even when going down to scales $p_T<2$ GeV.
Most of the cross section of Eq.~\ref{minijets} at fixed $p_T$ comes
from momentum fractions  $x \sim 2p_T/\sqrt s$,
so that a smaller scale $p_T$ also corresponds to a smaller $x$.
Therefore, when going towards smaller $p_T$,
there is a competition between the scale evolution
(causing a decrease in the parton density) and the
small-$x$ rise. As a result, the parton luminosity
does not decrease fast enough to cancel
the divergence in the $\hat t$- and $\hat u$-channels at small $p_T$.
This is basically the reason  for the rapid growth of the
non-screened cross sections as seen in Fig.~\ref{fig3}.

        In the top panel of Fig.~\ref{fig4} we show the average number of
produced minijets with $p_T\ge p_0$ and $|y|\le 0.5$ in a central
$AA$ collision with $A=200$. The curves are shown as functions of
$p_0$ and they are obtained simply by integrating the cross sections
in Fig.~\ref{fig3} over $p_T$ from $p_0$ upwards and  multiplying
by $T_{AA}({\bf 0})$. The average transverse energy carried by these
partons is obtained by weighting the cross sections of
Fig.~\ref{fig3} by $p_T$  and integrating again over $p_T$ in the
region $p_0\le p_T\le \sqrt s/2$. The corresponding  curves are
shown in the lower panel of Fig.~\ref{fig4} as functions of $p_0$.
Note also that $p_0$ here gives the time for formation of the system
according to Eq.~\ref{ftime}.

        We want to point out how the parameter $p_0$ is
needed as a cut-off only for the unscreened cross section.
This cut-off makes the unscreened minijet production finite
but the results depend strongly on the cut-off, as seen
in Fig.~\ref{fig4}. When  the self-consistent screening is included
the cross section saturates and there is no need for such a cut-off
anymore, provided that the saturation  occurs in the perturbative region
as it does in Fig.~\ref{fig3}. However, one may still ask how many
minijets are produced into the central rapidity unit with
transverse momentum larger than some lower limit $p_0$, and what is
the  average transverse energy from these quanta. Fig.~\ref{fig4}
gives an answer to these questions. Especially at the LHC energies,
where the  number of produced  minijets becomes very large, and the
parton  system very dense, the average transverse energy practically
saturates near  $p_0\sim 1$ GeV.

        The weaker $p_0$-dependence should also make the perturbative
energy density estimates of the early minijet system at
$\tau_f(p_0) = a/p_0$ more reliable.
In Fig.~\ref{fig5} the transverse energy density \cite{BJORKEN}
\begin{equation}
        \varepsilon = \frac{\bar E_T^{AA}(p_0)}{\pi R_A^2\tau_f(p_0)Y}
\end{equation}
is plotted against $p_0$.
Also the corresponding formation time is shown simultaneously.
It is interesting to notice how the maximum energy density
is reached at earlier times, and how the expansion of the
system starts to dominate  the evolution earlier
when screening is taken into account.  One should, however, keep
in mind that the finite overlap of the colliding nuclei will
reduce the energy density estimates especially at RHIC energies
\cite{KJEMG}. On the other hand,
roughly half of the released final transverse energy is expected to
originate from the soft particle production at RHIC energies
\cite{KLL,HIJING,KJEMG}.  This will slow down the decrease in
$\varepsilon$ at a later stage \cite{KJEMG}.

        The screening effect we discussed here is due to final
state interactions of produced partons. We should point out
that there will also be screening effects,
coming from initial state interactions or the color confining
radius in the dense parton clouds of a high energy nucleus \cite{mclerran}.
Ideally, one should take into account both screening effects in the
calculation of parton production. Nuclear shadowing due to
initial-state interactions will also reduce
the estimates on semihard parton production especially at the
LHC nuclear collisions, perhaps even at RHIC energies \cite{HIJING,KJE93}.
Nuclear gluon distributions are still uncertain to a large extent,
and especially the scale dependence of nuclear gluon shadowing
has to be studied carefully, e.g., as done  in \cite{KJE93}, but with
the modern parton distributions like used in this paper \cite{INPROG}.

        In summary, we have considered color screening of initial semihard
parton production in a phenomenological but self-consistent manner.
Following Ref.~\cite{BMW} we have computed the static electric
screening mass of the parton system  by using  the
number distributions of produced minijets and by taking into account
the formation time of the minijets. We do not try to estimate
the corresponding magnetic screening mass, instead we use the
obtained electric screening mass as a
regulator for the  divergent $t$ and $u$-channel sub-processes.
We have iteratively computed the lowest order
minijet cross section with a feedback from
the screening mass towards the region $p_T\sim \mu_D \lesssim 2$ GeV.
We have shown that for sufficiently large nuclei and for sufficiently
large energies, we obtain saturation of the minijet cross section
in the perturbative region $p_T\gg\Lambda_{QCD}$.
{}From this we conclude that the average number
of minijets in the central rapidity unit is finite and, furthermore,
there is no need for an {\it ad hoc} cut-off parameter $p_0$.
However, the question of how many minijets are produced with
$p_T\ge p_0$ is still quite valid, and we have shown that the answer
depends on the lower limit $p_0$ much more weakly than when
screening effects are neglected.
In the estimates of perturbative transverse-energy production
there is practically
no $p_0$-dependence below $p_0\lesssim 1$ GeV at the LHC energy.
Finally, we have shown how much the screening reduces  the Bjorken
estimate of the energy density of the early parton plasma.

\section*{Acknowledgements}

We thank the European Center for Theoretical Studies in Nuclear
Physics and Related Areas in Trento (Italy) for warm hospitality
and providing a stimulating environment for discussions.
We also thank M. Gyulassy, H. Heiselberg, K. Kajantie,
P.V. Ruus\-kanen and A. Sch\"afer for discussions.
KJE thanks the Academy of Finland for financial support.
This work was supported by the Director, Office of Energy
Research, Office of Nuclear Physics, Divisions of
Nuclear Physics of the U.S. Department
of Energy under Contract grant No.\ DE-AC03-76SF00098, DE-FG05-90ER40592.

\begin{figure}
  \caption[f1]{ An  example of a process contributing to the screening.
    Off-shell gluons are drawn in thick lines.
  \label{fig1} }
\end{figure}

\begin{figure}
\caption[f2]{
Longitudinally boost invariant space-time picture
of an $AA$ collision.  The $z$-axis is along the
beam, and the longitudinal proper time is $\tau=\sqrt{t^2-z^2}$.
Partons in the central region within a rapidity interval $Y$
(between the solid lines) and with transverse momenta $p_T> a/\tau_1$
have formed earlier than $\tau_1$, and thus screen the formation of
partons at $\tau\ge \tau_1$.
  \label{fig2} }
\end{figure}

\begin{figure}
  \caption[f3]{ ({\bf a}) Differential minijet cross section
$d\sigma_{\rm jet}/dp_Tdy$ at $y=0$ and screening mass $\mu_D$
as functions of transverse momentum $p_T$ in a $pp$ sub-process
of a central  $AA$ collision at $\sqrt s = 5.5\,A$TeV with $A=200$.
The dashed line is for unscreened cross section, the solid and
dotted curves for the screened case with different parameters $a$
from Eq. 1. Nuclear shadowing is not taken into account.
({\bf b}) The same as in panel (a) but for $\sqrt s=200\,A$GeV.
  \label{fig3} }
\end{figure}

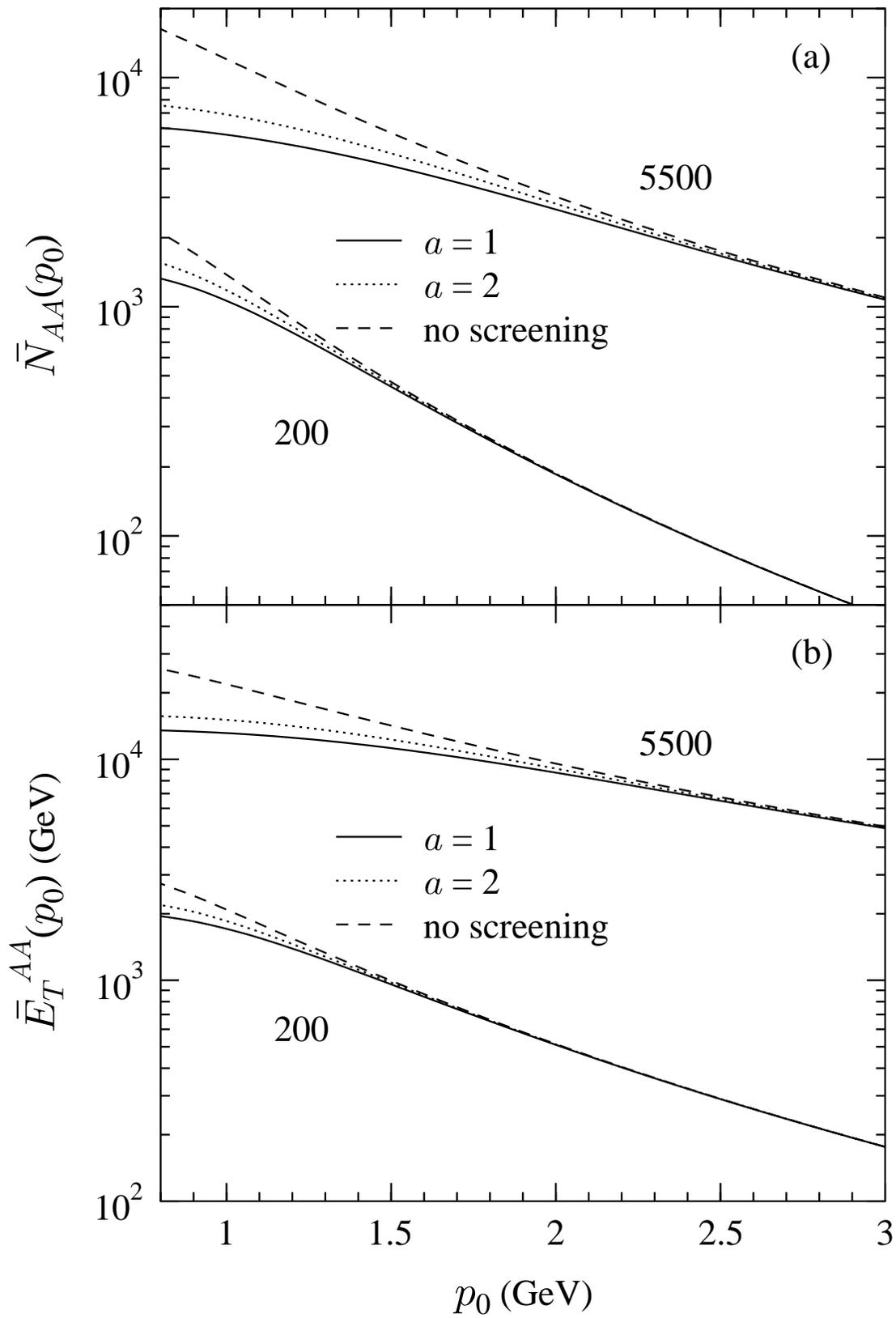
\begin{figure}
  \caption[f4]{  ({\bf a}) The average number of minijets produced with
$p_T\ge p_0$ and $|y|\le 0.5$ in a central $AA$ collision as a
function of $p_0$ (see the text). The solid, dotted and dashed
curves correspond to the curves in Fig. 3, and the two sets of
curves are for energies $\sqrt s=200$ and 5500 $A$GeV.
({\bf b}) The average transverse energy of the minijets of panel (a)
as a function of $p_0$ (see the text). Labeling of the curves is the
same as in panel (a).
  \label{fig4} }
\end{figure}

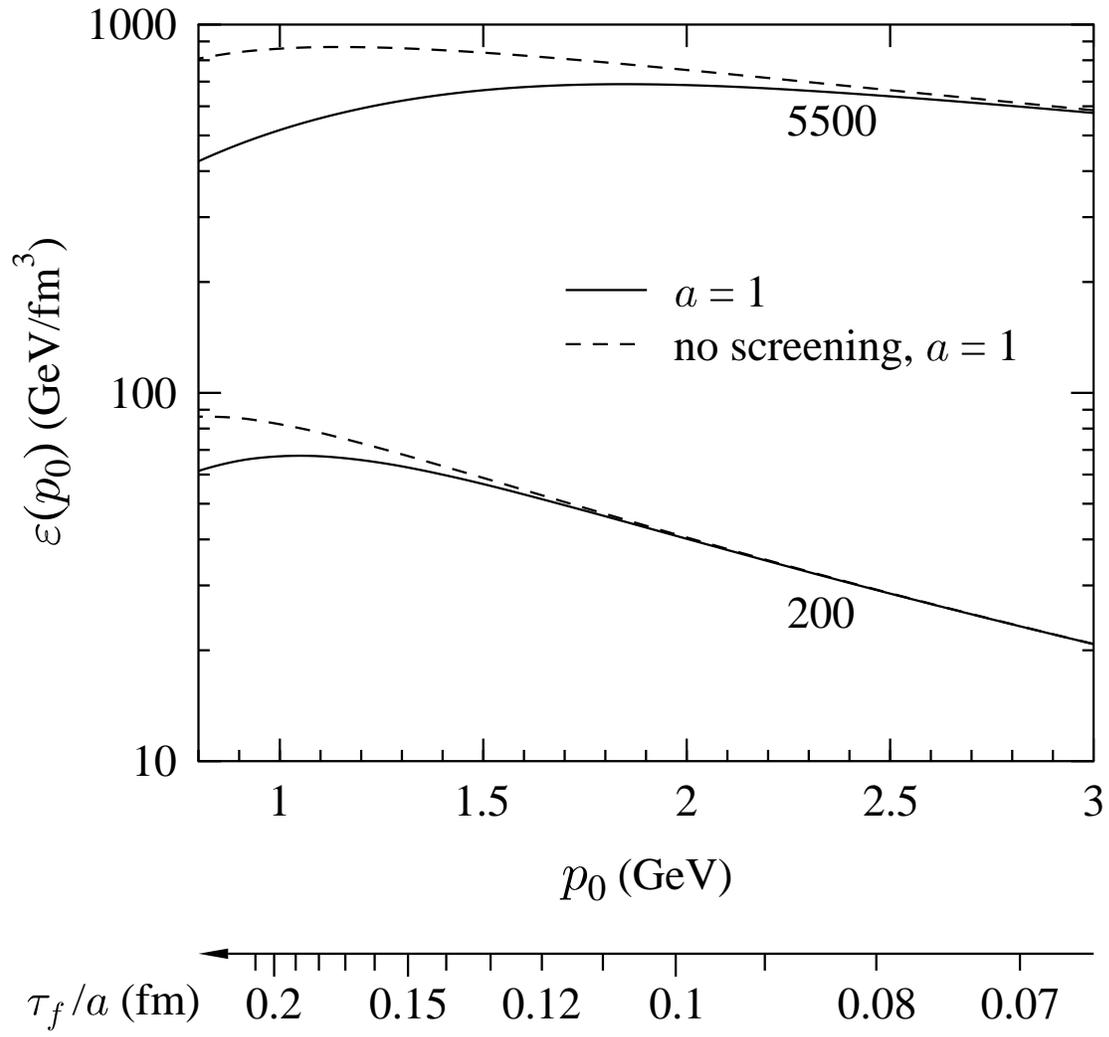
\begin{figure}
\caption[f5]{ Transverse energy density $\varepsilon$ of the
perturbatively produced minijet system as a function of the
smallest transverse momentum $p_0$. The corresponding
formation time axis is also shown. The upper set of curves
is for $\sqrt s = 5500\, A$GeV and the lower set for
$\sqrt s = 200\, A$GeV.
The solid and dashed lines show the estimate with and without
screening, respectively, for $a=1$.
  \label{fig5} }
\end{figure}

\end{document}